# INTERACTION OF NON-RELATIVISTIC PARTICLE WITH SPIN WITH A PSEUDOSCALAR FIELD

Shurgaia

I. Javakhishvili State University, A. Razmadze Mathematical Institute

1 M. Alexidze str. 0171 Tbilisi, Georgia

A non-relativistic model of a moving fermion in a pseudoscalar field is investigated using N. N. Bogolyubov's method. The invariance properties are rigorously described. The energies of the ground and excited states of the system are obtained.

In the proposed work, based on N. N. Bogolyubov's canonical transformations [1, 2], the interaction of a non-relativistic spin particle with a neutral pseudoscalar field is studied. In the model under study, similar to work [3], where the problem with a spinless particle is considered, oscillatory levels naturally arise as a result of the strong interaction of the particle with the field. In addition, the effects associated with the presence of spin in the particle are described. It is shown that, in comparison with the fixed-source model, a splitting of the energy levels occurs. The model under study (but with a non-gradient connection) was studied by a different method in [4]. The isobar levels obtained coincide with a similar result to the fixed-source models. This is due to the fact that in [4] the conservation laws are violated. A rigorous analysis of the invariance properties shows that the energy levels are split relative to the magnetic quantum number.

## Model

Let the system be enclosed in a cube of volume V and the Hamiltonian be written as

$$H = \frac{p^2}{2m} + \frac{1}{2}\sum_f \omega_f(a_f^+ a_f + a_f a_f^+) + \frac{g}{\sqrt{2}}\sum_f B_f(f\sigma)(a_f + a_{-f}^+)e^{ifX} \quad (1)$$

where $a_f$ are Bose field operators, $\boldsymbol{p} = -i\nabla_X$ is the particle momentum, and $m$ is its mass. The coefficient $B_f$ is related to the source function normalized to unity by equality

$$B_f = (V\omega_f\mu^2)^{-1/2} \int d\boldsymbol{x}\rho(\boldsymbol{x})e^{-if x}. \quad (2)$$

Here $\boldsymbol{p} = -i\nabla_X$ the momentum operator of the particle, $\boldsymbol{\sigma}$ Pauli matrix. The Hamiltonian $H$ is invariant with respect to transformations from the group of motion of the three-dimensional Euclidean space $M(3)$, i.e. with respect to transformations

$$x_\alpha \to C_{\alpha\nu}x_\alpha + q_\alpha, \quad a_f \to a_{f'}t_{f'f}, \quad a_f^+ \to t_{ff'}a_{f'}^+, \quad (3)$$

where $q_\alpha$ are the parameters of the 3-dimensional shift, $C_{\alpha\nu}$ is the rotation matrix parametrized Euler angles $\varphi, \eta, \psi$. The matrix $t_{ff'}$ is the representations of the group in the basis of exponential functions, and also $\sum_{f'} \bar{t}_{ff'} t_{f'l} = \delta_{fg}$. The invariants of the group are the square of the generator of the 3-dimensional translation and the projection of the vector of the rotation generators onto the translation operator. Therefore, the development of the system occurs according to the laws of conservation of momentum and the projection of the momentum onto the direction of the momentum. In what follows, it is convenient to introduce complex coordinates and momenta of the field:

$$q_f = \frac{a_f + a_{-f}^+}{g\sqrt{2}}, \quad p_f = g\frac{a_f^+ - a_{-f}}{g\sqrt{2}}. \tag{4}$$

The Hamiltonian, momentum and angular momentum of the system are now determined by the equalities:

$$H = \frac{p^2}{2m} + \frac{1}{2}\sum_f \omega_f \left(g^2 q_f^+ q_f + \frac{p_f p_f^+}{g^2}\right) + g^2 \sum_f B_f(f\sigma) q_f e^{ifX} \tag{5}$$

$$P_\alpha = p_\alpha - \frac{i}{2}\sum_f \{q_f p_f + p_f q_f\} \tag{6}$$

$$L_\alpha = \frac{1}{2}\sigma_\alpha + \varepsilon_{\alpha\beta\gamma} X_\alpha p_\gamma - \frac{i}{2}\sum_f \{p_f(\hat{l}_\alpha q_f) + (\hat{l}_\alpha q_f) p_f\} \tag{7}$$

Here $\hat{l}_\alpha$ are rotation generators of the symmetry group $M(3)$.

Let's introduce new variables instead of $X_\alpha$ and $q_f$:

$$X_\alpha = \frac{1}{\sqrt{g}} C_{\alpha\beta}(\varphi, \vartheta, \psi)\lambda_\beta + q_\alpha, \quad q_f = t_{ff'}(\mathbf{q}, C)\left(u_{f'} + \frac{1}{g} Q_{f'}\right). \tag{8}$$

Three Euler angles $\varphi, \vartheta, \psi$ and the 3-dimensional vector $q_\alpha$ together with $\lambda_\alpha$ and $Q_f$ constitute a set of new independent variables. The number of variables has increased by six. To preserve the total number of independent variables, we impose the same number of additional conditions on the variables $Q_f$:

$$\sum_f N_{\alpha f}^i Q_f = 0, \quad \text{for } i = 1,2, \ \alpha = 1,2,3. \tag{9}$$

where the index $i$ is used to label the subgroups of translation for $i = 1$ and rotation for $i = 2$. There exists a function $M_{\alpha f}^i$ obeying equality:

$$\sum_f N_{\alpha f}^i M_{f\beta}^k = \delta_{ik}\delta_{\alpha\beta}. \tag{10}$$

The impulses are calculated according to the scheme developed in [2]. When differentiating, it is convenient to represent the matrix U as a product of the displacement and then rotation matrices

$$t_{ff'}(\mathbf{q}, C) = \sum_l t_{ff'}(\mathbf{q}, l)\, t_{ff'}(\mathbf{0}, C).$$

To take into account additional conditions (9), a projection matrix is defined $A_{ff'} = \delta_{ff'} - M^i_{f\beta} N^i_{\alpha f'}$ with the properties:

$$\sum_l A_{fl} M^i_{l\beta} = \sum_l N^i_{\alpha l} A_{lf} = 0. \tag{11}$$

With its help, the variables $Q_f$ are linearly expressed through the independent variables $Q_f = \sum_l A_{fl} z_l$ and conditions (9) are satisfied automatically due to the properties (11). The calculation is carried out under the assumption that $q_f$ but not $X_\alpha$, depend on the parameters of the symmetry group. Taking into account the above, we can obtain the following expression for $p_\alpha$ and $p_f$:

$$p_\alpha = C_{\alpha\beta} \lambda_\beta, \tag{12}$$

$$p_f = \sum_{f'} t_{ff'} \left\{ g P_{f'} - i\tilde{N}^1_{\sigma f'}\left(\bar{l}^1_\sigma - \varepsilon_{\sigma\nu\alpha}\lambda_\nu p_{l_\alpha} + i\sum_{kk'}(\bar{J}^1_\sigma)_{kk'} Q_{k'} P_k\right) - i\tilde{N}^2_{\sigma f'}\left(\bar{l}^2_\sigma - \sqrt{g} p_{\lambda_\sigma} + i\sum_{kk'}(\bar{J}^2_\sigma)_{kk'} Q_{k'} P_k\right) \right\} \tag{13}$$

where $P_f = \sum_l A_{lf}(-i\,\partial/\partial Q_l)$ and satisfy constraint similar to (9).

$$\sum_f M^i_{f\alpha} P_f = 0, \tag{14}$$

The operators $\bar{l}^1_\sigma$ and $\bar{l}^2_\sigma$ are differential forms depending on the parameters of the group, and are essentially infinitesimal operators of the inverse group M(3), and the matrices $\bar{J}^l_\sigma$ are infinitesimal matrices of the algebra of the same group. The quantities $\tilde{N}^2_{\sigma f}$ satisfy the algebraic equations:

$$\tilde{N}^i_{\alpha f} = -N^i_{\alpha f} - \frac{1}{g}\sum_l N^i_{\alpha l}(\bar{J}^j_\sigma)_{ll'} Q_{l'} \tilde{N}^j_{\sigma f}, \tag{15}$$

containing a small parameter $g^{-1}$ thanks to which equations (15) can be solved by the iteration method. It is necessary to introduce another transformation affecting the spin variables of the particle:

$$S^{-1} \sigma_\alpha S = C_{\alpha\beta} \sigma_\beta \tag{16}$$

and leading to the transformation of the wave function and the Hamiltonian of the system according to the relations

$$H \to H' = S^{-1} H S, \quad \Psi \to \Psi' = S^{-1} \Psi.$$

The explicit form of the unitary operator $S$ is well known from the theory of the group $SU(2)$. This transformation affects the operator $\bar{l}_\alpha^1$, but not $\bar{l}_\sigma^2$:

$$S^{-1}\bar{l}_\alpha^1 S = \bar{l}_\alpha^1 - \frac{1}{2}\sigma_a. \tag{17}$$

It remains to express the operators of momentum and momentum of the system in the new representation. It's easy to show that after the introduced transformations we obtain:

$$P_\alpha = l_\alpha^1, \qquad S^{-1}L_\alpha S = l_\alpha^1. \tag{18}$$

The operators $l_\alpha^1$ and $l_\alpha^2$ are infinitesimal operators of the group M(3) and satisfy the usual commutation relations for moments and momenta. At the same time, they commute with the operators $\bar{l}_\alpha^1$ and $\bar{l}_\alpha^2$. Further, since the potential energy of the system, defined by equality

$$H_{\text{pot}} = g^2 \sum_f B_f B(f\sigma) \exp\left(\frac{i}{\sqrt{g}}f\lambda\right)\left(u_f + \frac{1}{g}Q_f\right) +$$

$$+\frac{1}{2}\Sigma_f\{u_f^* u_f + u_f^* Q_f + Q_f^+ u_f + Q_f^+ Q_f + Q_f^+ u_f + Q_f^+ Q_f\} \tag{19}$$

also does not depend on the group parameters it commutes with the operators $l_\alpha^1$, $l_\alpha^2$. Having written out the kinetic energy, we can verify that its dependence on the parameters is determined only by the $\bar{l}_\alpha^i$ operators. This fact clearly indicates the exact fulfillment of the conservation laws and allows us to isolate the dependence of the wave function on the parameters of the symmetry group in the form of a factor

$$\psi' = t(Q, C)\Psi''(\lambda_\alpha, \sigma_\alpha, Q_f). \tag{20}$$

and replace the operators $\bar{l}_\alpha^1$, $\bar{l}_\alpha^2$ by c-numbers

$$\bar{l}_\alpha^1 \to \bar{J}_\alpha, \quad \bar{l}_\alpha^2 \to P_\alpha$$

Computation of energy spectrum of the system will now be strictly consistent with conservation laws.

2. Let us proceed to the expansion of the Hamiltonian in a series of inverse powers of g. First, let us make several useful transformations. Let us introduce, similarly to [3], the quantity $I_\alpha$, related to the momentum $P_\alpha$ by the equality

$$P_\alpha = g^2 I_\alpha. \tag{21}$$

This leads to an increase in the order of momentum. As a result, translation effects can be seen already in the first approximation. Next, it is necessary to introduce another transformation of the wave function:

$$\Psi'' \to \exp\left(ig \sum_f s_f Q_f\right) \Psi''(q, C, Q_f)$$

which means for the momentum $P_f$ following replacement: $P_f \to g s_f + P_f$. The c-numbers $s_f$ obey equalities

$$\sum_f M_{f\alpha}^i s_f = 0. \tag{22}$$

Now we can begin to expand the Hamiltonian. To do this, we should substitute the numbers Natl into the kinetic energy of the field in the form of an iterative series in powers of $g^{-1}$. In addition, we will expand in a series in fractional powers of $g$ $\exp\left(\frac{i}{\sqrt{g}} f\lambda\right)$ preserving only the first three terms:

$$\exp\left(\frac{i}{\sqrt{g}} f\lambda\right) = 1 + \frac{i}{\sqrt{g}} f\lambda - \frac{1}{2g} f\lambda^2 \ldots$$

We have to expand in series in powers of $g^{-1}$ the energy and the wave function $\Psi''$ also:

$$E = g^2 E_0 + g E_1 + E_2 + g^{-1} E_3 + \cdots$$
$$\Psi = \Psi_0 + g^{-1} \Psi_1 + \cdots \tag{23}$$

Substituting these expansions into the Schrödinger equation, one can develop perturbation theory in inverse powers of g. The zeroth approximation contains term$\sigma$ of order $g^2$:

$$H_0 = \sum_f B_f(\boldsymbol{f\sigma}) u_f + \frac{1}{2} \sum_f \omega_f \left(|u_f|^2 + |\alpha_f|^2\right)$$

(24)

with

$$\alpha_f = s_f + i N_{\alpha f}^i I_\alpha. \tag{25}$$

In this approximation, the spin variables are separated from the others: $\Psi_0 = \chi_0 \Phi_0(\lambda_\alpha Q_f)$. The function $\chi_0$ satisfies the equation $H_0 - E_{0)}) \chi_0 = 0$. For definiteness, directing the vector $\boldsymbol{a} = \sum_f B_f u_f \boldsymbol{f}$ along the z-axis, we can obtain the following solutions to the equation:

$$E_0^{(1)} = -a_3 + \frac{1}{2} \sum_f \omega_f \left(|u_f|^2 + |\alpha_f|^2\right) \quad \chi_0^1 = \begin{pmatrix} 0 \\ 1 \end{pmatrix} \tag{26}$$

$$E_0^{(2)} = a_3 + \frac{1}{2} \sum_f \omega_f (|u_f|^2 + |\alpha_f|^2) \quad \chi_0^1 = \begin{pmatrix} 1 \\ 0 \end{pmatrix} \tag{27}$$

Next, we need to find higher-order corrections in the expansion parameter to E^ as the energy of the ground state. Having written out the terms of order g, we see that the variables $\lambda_\alpha$ and $Q_f$ are separated: $\Phi_0^{(1)} = F(\lambda)\Theta(Q_f)$. In this approximation, we can determine the dependence of the wave function on $\lambda$ and, based on the regularity of the function $\Theta(Q_f)$ in the variables $Q_f$ the values of the numbers $u_f$ and $\alpha_f$. Omitting detailed calculations, since they repeat the work [3], we write out the value of the numbers $\alpha_f^*$ at and и(:

$$\alpha_f^* \omega_f = -iM_{fa}^2 c_a, \quad u_f = \frac{\omega_f B_f f_3}{\omega_f^2 - (fc)^2}. \tag{28}$$

where $c$ is the vector of the average velocity of the particle. For the value $I_\alpha$, we can obtain

$$I_\alpha = \sum_f \frac{f_\alpha(fc)}{\omega_f} |u_f|^2. \tag{29}$$

Note that the terms of order $g^{3/2}$ vanish.

Continuing the analogy with work [3], we note that the equation according to Xa describes the oscillations of a particle. However, the states corresponding to them cannot be considered stationary. To construct such, it is necessary to include in the Hamiltonian of this approximation the dipole interaction of the particle with the field (approximately of order $\sqrt{g}$) and the bilinear form in $Q_f$ and $P_f$ (approximately of order g°). It is convenient to represent the wave function in the form

$$\Phi_0^{(1)} = \exp\left(\frac{i}{\sqrt{g}} mc\lambda\right) \widetilde{\Phi}_0^{(1)}(\lambda_\alpha, Q_f).$$

This representation removes from the terms of order $\sqrt{g}$ the terms linear in $-\partial/\partial\lambda_\alpha$. The equation obtained in this way is diagonalized using the known canonical transformations [1, 3]. However, we will not dwell on this, noting that after the transformation the function $\widetilde{\Phi}_0^{(1)}(\lambda_\alpha, Q_f)$ is defined in the space of occupation numbers. The corresponding energies, the energies of the oscillators, correspond to the stationary states of the system.

3. To study the internal structure of the system associated with the presence of spin in a particle, it is necessary to study the terms of the Hamiltonian up to order $g^{-2}$. By combining together the terms of order $g^{-1/2}, g^{-1}$ and denoting them by $H_3$, it can be shown that the energy in this approximation is equal to

$$E_3 = \frac{3}{4}K^2\sqrt{m/\gamma} \tag{30}$$

in which

$$K = \frac{1}{3}\sum_f \omega_f f^2 |v_f|^2, \quad \gamma = \frac{1}{3}\sum_f \frac{|B_f|^2 f^2 f_3^2}{\omega_f^2 - (fc)^2}.$$

The contribution of the terms of order $g^{-3/2}$ vanishes. The condition of solvability of the equation in the approximation $g^{-2}$ allows us to calculate the energy $E_4$ in this order. We do not give the explicit form of equation due to their cumbersomeness. We only note that the main interest is in the terms containing the quantities $J_\alpha, \sigma_\alpha$. The remaining terms, after averaging over the ground state function, make an additive contribution to the energy $E_4$ and do not change its characteristic structure. For simplicity, we omit them. The final expression of $E_4$ with this comment is:

$$E_4 = \frac{N^2}{2}\left\{j(j+1) + m + \frac{3}{4}\right\}. \tag{31}$$

where $N^2 = \frac{1}{3}\sum_f \omega_f |f \times \nabla_f v_f|^2$. The quantum number $j = {}^1\!/_2, {}^3\!/_2, {}^5\!/_2 \ldots$ corresponds to the total angular momentum of the system. The number $m$ takes on values of the z-component of the angular momentum: $-j \leq m \leq j$.

4. Let us briefly summarize the obtained results. We have constructed a scheme of successive approximations to the energy and wave function of the system, strictly taking into account the law of conservation of momentum. We have obtained a picture of the interaction, which is developed in the model of a spinless nonrelativistic particle [3]. An expression for the momentum is given, which coincides with the similar expression in [3]. It is found that the energy of the ground state of the system is determined by the expression.

$$E_0 = -g^2 \sum_f u_f B_f f_3 + \frac{1}{2}g^2 \sum_f |u_f|^2 \left(\omega_f + \frac{(fc)^2}{\omega_f}\right) - \frac{1}{2}mc^2 + g\varepsilon_0 \tag{32}$$

where $\varepsilon_0$ is the minimum eigenvalue of the Hamiltonian of order g taking into account the dipole interaction of the particle with the field. Considering the limit of a fixed source, which corresponds to $c = 0$, we find that $E_0$ goes over into the static eigenenergy of the source.

Let us turn to formula (32), which defines the excited states of the system. In contrast to the theory of a fixed source, a splitting of the levels occurs. The number of such levels is $2j + 1$. Indeed, in fixed source theories all directions are equal (the particle is at rest), and, consequently, the spectrum is degenerate with respect to the magnetic quantum number. In the model under study, the distinguished direction is the direction of motion. In this case, as is known, the role of a good quantum number is played by the projection of the angular momentum onto the momentum. With respect to it, the energy in (32) is degenerate, but the levels are split with respect to the magnetic quantum number $m$. In the fixed-source limit, $m = -1/2$ [5] and

$$E_4 = \frac{N^2}{2}\left\{j(j+1) + \frac{1}{4}\right\}$$

which coincides with the corresponding result of fixed-source models. Indeed, in this case, as is well known, quanta with unit angular momentum participate in the interaction. They can be additively isolated by expanding the field operators in spherical harmonics. The number of possible states of such quanta is three. After the canonical transformation, the two Euler angles and the projection of the coordinates of the mesons interacting with the source onto the z axis are taken as independent variables. To implement such a transition in our model, we must set the angle "f equal to zero. This means that the operator $\bar{l}_\alpha$ is fixed in the plane (1, 2), i.e. the eigenvalue of the operator $\bar{l}_3$ is equal to zero. Therefore, after the transformation using the operator $S$, the eigenvalue $m$ of the operator $\bar{l}_3$—the projection of the total momentum of the system— is equal to —1/2, i.e. the eigenvalue of the spin in the ground state.

# Appendix

We will give some necessary formulae.

The matrix elements of the presentation of the group $M(3)$ in the basis of exponential functions

$$t_{ff'} = e^{-ifq}\delta(f - f').$$

Infinitesimal matrices of algebra:

$$(J_\alpha^1)_{ff'} = -i\varepsilon_{\alpha\beta\gamma}f_\beta \frac{\partial}{\partial f_\gamma}\delta(f - f').$$

$$(J_\alpha^1)_{ff'} = f_\alpha \delta(f - f').$$

Generators of the group $M(3)$:

$$l_\alpha^1 = j_\alpha - i\varepsilon_{\alpha\beta\gamma}q_\beta \frac{\partial}{\partial q_\gamma}, \quad l_\alpha^2 = -i\frac{\partial}{\partial q_\alpha}$$

where $j_\alpha$ are well known generators of the group $SO(3)$.

Generators of inverse transformation from the group $M(3)$:

$$\bar{l}_\alpha^1 = \bar{J}_\alpha, \quad \bar{l}_\alpha^2 = \bar{C}_{\alpha\beta}\left(-i\frac{\partial}{\partial q_\beta}\right).$$

The numbers $M_{fa}^i$ and $N_{fa}^i$ are chosen as follows;

$$N_{fa}^i = \sum_l (J_\alpha^i)_{fl} u_l, \quad M_{fa}^i \sum_l (J_\alpha^i)_{lf} v_l$$